\documentclass[usenatbib]{mn2e}
\usepackage{graphicx}

\newcommand\op[1]{{#1}}

\newcommand\un[1]{{\,\rm #1}}
\newcommand\rs[1]{_\mathrm{#1}}
\newcommand\g{$\gamma$}
\newcommand\ApJ{ApJ}

\title[Synchrotron and IC images of SNRs]{Some properties of synchrotron radio and inverse-Compton gamma-ray images of supernova remnants}
\author[Petruk O. et al.]{O.~Petruk$^{1,2,4}$, V.~Beshley$^{2}$, F.~Bocchino$^{3,4}$, S.~Orlando$^{3,4}$\\
$^{1}$Institute for Applied Problems in Mechanics and Mathematics, Naukova St.\ 3-b,
   79060 Lviv, Ukraine\\
$^{2}$Astronomical Observatory, National University, Kyryla and Methodia St.\ 8, 79008 Lviv, Ukraine\\
$^{3}$INAF - Osservatorio Astronomico di Palermo ``G.S.
              Vaiana'', Piazza del Parlamento 1, 90134 Palermo, Italy\\
$^{4}$Consorzio COMETA, via Santa Sofia 64, 95123 Catania, Italy
}

\begin{document}

\date{Accepted .... Received ...; in original form ...}

\pagerange{\pageref{firstpage}--\pageref{lastpage}} \pubyear{2008}

\maketitle

\label{firstpage}

\begin{abstract}
The synchrotron radio maps of supernova remnants (SNRs) 
in uniform interstellar medium and interstellar magnetic field (ISMF) are analysed, 
allowing different `sensitivity' of injection efficiency to the shock obliquity. 
The very-high energy \g-ray maps due to inverse Compton process  
are also synthesized. 
The properties of images in these different wavelength bands are compared, 
with particulr emphasis on the location of the bright limbs in bilateral SNRs. 
Recent H.E.S.S. observations of SN~1006 show that the radio and IC \g-ray limbs 
coincide, and we found that this may happen if: 
i) injection is isotropic but the variation of the maximum energy of electrons is 
rather quick to compensate for differences in magnetic field; 
ii) obliquity dependence of injection (either quasi-parallel or 
quasi-perpendicular) and the electron maximum energy is strong enough 
to dominate magnetic field variation.  
In the latter case, the obliquity dependences of 
the injection and the maximum energy should not be opposite. 
We argue that the position of the limbs alone and even their coincidence in radio, 
X-rays and \g-rays, as it is discovered by H.E.S.S. in SN~1006, cannot be 
conclusive about the dependence of the 
electron injection efficiency, the compression/amplification of ISMF and 
the electron maximum energy on the obliquity angle.
\end{abstract}


%%%%%%%%%%%%%%%%%%%%%%%%%%%%%%%%%%%%%%%%%%%%%%%%%%%%%%%%%%%%%%%%%%%%%%%%%
\section{Introduction}

The observation of the supernova remnants (SNRs) in very-high energy (VHE) \g-rays by H.E.S.S. 
and MAGIC experiments is an important step toward understanding the nature of the 
Galactic cosmic rays and kinematics of charged particles and magnetic field in vicinity of the strong 
nonrelativistic shocks. 
However, the spectral analysis of multi-wavelenght data allows both 
for leptonic and hadronic origin 
of VHE \g-ray emission (e.g. RX J1713.7-3946: \citet{RX1713Ber-Volk-06}, 
\citet{RX1713aha2007}). 
In this context, the broad-band fitting of the spectrum of the nonthermal emission from SNRs 
is one of the hot topics in present studies of SNRs. At the same time, another very important 
source of scientific information, the distribution of the surface brightness, is not in great demand. 
There are just some discussions emphasyzing that observed correlations of brightness in radio, 
X-rays and \g-rays may be considered to favor electrons to be responsible for VHE emission in 
RX~J1713.7-3946, Vela Jr. and some other SNRs (e.g. \citet{RX1713aha2006}, 
\citet{Plaga2008}). However, should the patterns of surface brightness in radio, X-rays and 
\g-rays realy correlate if the VHE \g-radiation originates from electrons? 
What should be the limitations for theory once observed patterns are really quite similar, 
especially in symmetrical bilateral SNRs, like in SN~1006 (H.E.S.S. Source of the Month, August 2008).  

Another key issue for particle kinetics is the 3-D morphology of bilateral SNRs in general and  
SN~1006 particularly. Is it polar-cap or barrel-like? The answer of this question is strongly related 
to the model of injection (quasi-parallel in the former and isotropic or quasi-perpendicular in the 
latter case), giving therefore an important hint for acceleration theory. The properties of 
brightness distribution may be the most conclusive issue in this task (e.g. criterion of 
\citet{Rotetal04}, azimuthal profiles comparison in \citet{pet-SN1006mf}). 

An experimental investigation of SNR images have to be complemented with theoretical modelling of SNR maps in different energy domains. Radio and X-ray synchrotron images in the uniform interstellar medium (ISM) and the uniform interstellar
magnetic field (ISMF) are modeled by \citet{Reyn-98}. The role of gradient of ISM density and ISMF strength on radio morphology of SNRs are studied by 
 \citet{Orletal07}. These papers bases on the classical MHD and assumes unmodified shocks. 
Studies on nonthermal images of SNRs with non-linear acceleration theory undergo development \citep{Ell2008-images}. The profiles of the synchrotron brightness in such SNRs are subject of investigation in 
\citet{Ell-Cassam2005-profiles} and \citet{Decours-2005-prof}. 

In the present paper, we present for the first time the inverse-Compton \g-ray images of SNRs in uniform ISM and ISMF produced on the basis of the model of \citet{Reyn-98}. In addition to this model, we allow for different `sensitivity' of injection efficiency to the shock obliquity like it is apparent in numerical results of \citet{ell-bar-jones-95}. The synthesized maps are compared with the radio ones. Some consequencies for origin of VHE emission of SNRs and electron injection scenario are drawn.

%%%%%%%%%%%%%%%%%%%%%%%%%%%%%%%%%%%%%%%%%%%%%%%%%%%%%%%%%%%%%%%%%%%%%%%%%
\section{Model}

We consider SNR in uniform ISM and uniform ISMF. 
At the shock, the energy spectrum of electrons is taken as 
$N(E) = KE^{-s}\exp\left(-E/E\rs{max}\right)$, 
$E\rs{max}$ is the maximum energy of electrons, 
$s=2$ is used throughout of this paper. 
We follow \citet{Reyn-98} in calculation of the 
evolution of the magnetic field and relativistic electrons 
(see details also in \citet{petruk2006}, \citet{Pet-Beshl-en-2008}). 
The compression 
factor for ISMF $\sigma\rs{B}$ increases from unity 
at parallel shock to 4 at perpendicular one.
The fiducial energy at parallel shock,  
which is responsible for the `sensitivity' of relativistic electrons 
to the radiative losses \citep{Reyn-98} 
and which is used in IC images is set to $E\rs{max}$. 
The synchrotron losses are considered as the dominant channel for the 
radiative losses of relativistic electrons. 
We assume that $K$ is constant in time; eventual evolution of 
$K$ affects the radial thickness of rims and does not modify the main features of 
the surface brightness pattern \citep{Reyn-98}. 

\op{Electrons emitting IC photons have energies $E\sim E\rs{max}$. 
Like $K$, $E\rs{max}$ is assumed to be constant in time. 
Its possible variation in time does not change the pattern of IC brightness 
and leads to effects similar to those originating from the time dependence of $K$. 
Namely, features in IC images have to be radially thicker if 
$E\rs{max}$ decreases with time (i.e. increases with the shock velocity): 
since $E\rs{max}$ was larger at previous times, 
there are more electrons in the SNR interior able to emit IC photons at the present time. 
If $E\rs{max}$ increases with time (i.e. decreases with the shock velocity) then 
maxima in brightness are expected to be radially thinner. 
}

%--------------------------------------------------------------
\begin{figure}
 \centering
 \includegraphics[width=8.0truecm]{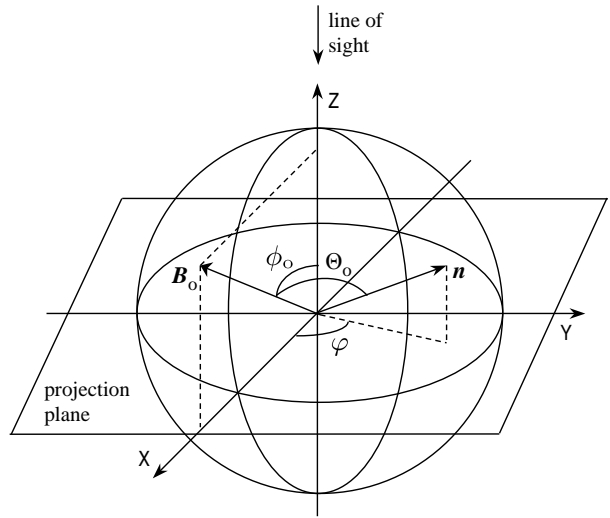}
 \caption{Geometry of the task. The obliquity angle $\Theta\rs{o}$, 
 the aspect angle $\phi\rs{o}$ and the azimuth angle $\varphi$ are shown. 
 ISMF $B\rs{o}$ is chosen to be parallel to the X0Z plane. 
                }
 \label{thetaK:angles}
\end{figure}
%--------------------------------------------------------------
%--------------------------------------------------------------
\begin{figure*}
 \centering
 \includegraphics[width=12.9truecm]{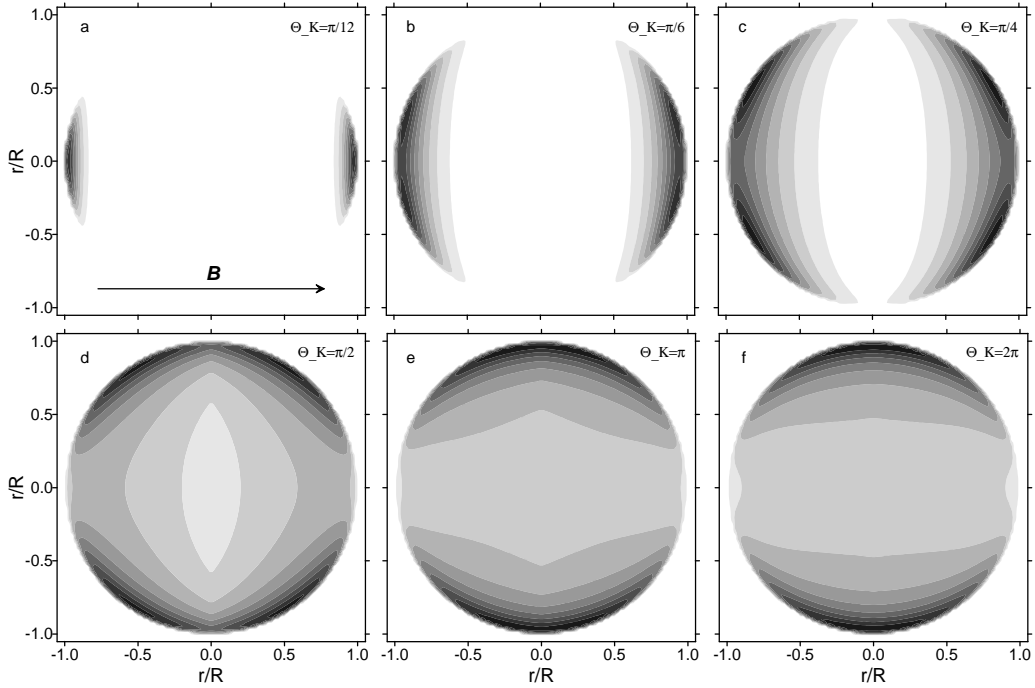}
 \caption{Radio images of SNR for an aspect angle $\phi\rs{o}=90^\mathrm{o}$ 
 and different $\Theta\rs{K}$: 
 $\pi/12$ (a), $\pi/6$ (b), $\pi/4$ (c), 
 $\pi/2$ (d), $\pi$ (e), $2\pi$ (f). 
 Ambient magnetic field is oriented along the horizontal axis. 
 Hereafter, the increment in brightness is $\Delta S=0.1 S\rs{\max}$.
                }
 \label{thetaK:fig1}
\end{figure*}
%--------------------------------------------------------------

\citet{Reyn-98} considered three models for injection: quasi-parallel, 
isotropic and quasi-perpendicular. The pattern of the radio surface 
brightness distribution in the case of the quasi-perpendicular injection is quite similar 
to the isotropic injection case, though with different contrasts 
\citep{reyn-fulbr-90,Orletal07}. 
The numerical calculations of \citet{ell-bar-jones-95} show that 
the obliquity dependence of the injection efficiency $\varsigma$ (a fraction of 
accelereted electrons) may be either flatter or steeper than 
in the classic quasi-parallel case ($\varsigma\propto \cos^2\Theta\rs{o}$ where $\Theta\rs{o}$ 
is the obliquity angle, the angle between the ISMF and the normal to the shock, Fig.~\ref{thetaK:angles}). 
In order to be more general than \citet{Reyn-98}, we allow the injection efficiency 
to vary with obliquity angle with different `sensitivity' which is given 
by the parameter $\Theta\rs{K}$:
\begin{equation}
 \varsigma(\Theta\rs{o})=\varsigma_{\|}
 \exp\left(-\big({\Theta\rs{o}}/{\Theta\rs{K}}\big)^2\right)
 \label{finj}
\end{equation}
where $\varsigma_{\|}$ is the efficiency for the parallel shock. 
This expression restores approximately 
the results of \citet{ell-bar-jones-95}
with $\Theta\rs{K}=\pi/9\div \pi/4$. 
The classic quasi-parallel injection may be approximated with $\Theta\rs{K}=\pi/6$. 
Isotropical injection assumes $\Theta\rs{K}=\infty$, but the values 
$\Theta\rs{K}\geq 2\pi$ produces almost the same images as $\Theta\rs{K}=\infty$ 
because the range for obliquity angle is $0\leq \Theta\rs{o}\leq \pi/2$. 

We consider also {quasi-perpendicular} injection: 
\begin{equation}
 \varsigma(\Theta\rs{o})=\varsigma_{\|}
 \exp\left(-\big({(\Theta\rs{o}-\pi/2)}/{\Theta\rs{K}}\big)^2\right).
 \label{finjperp}
\end{equation}
%where $\varsigma_{\bot}$ is the efficiency for the perpendicular shock. 

\op{
In the most cases presented here, $E\rs{max}$ is assumed to be constant over SNR surface; 
this choice allows us to clearly see the role of other parameters. 
\citet{Reyn-98} considered loss-limited, time-limited and escape-limited models for $E\rs{max}$. 
In all cases, except of the loss-limited one with the level of turbulence comparable with the Bohm limit, 
$E\rs{max}$ should grow with increase of $\Theta\rs{o}$ \citep{Reyn-98}. 
We model the role of possible increase of $E\rs{max}$ with obliquity with a simple 
parameterization 
\begin{equation}
 E\rs{max}(\Theta\rs{o})= E\rs{max\|} 
% A\rs{E} \sin^2\left(\Theta\rs{o}\right)
 \exp\left(-\big({(\Theta\rs{o}-\pi/2)}/{\Theta\rs{E}}\big)^2\right)
 \label{fEmaxperp}
\end{equation}
where $\Theta\rs{E}$ is a parameter, $E\rs{max\|}$ the maximum energy at parallel shock. 
This formula, with different values of $\Theta\rs{E}$, is able to restore approximately 
different cases considered by \citet{Reyn-98}. 
}

The surface brightness is calculated integrating emissivities along the line 
of sight within SNR. 
The synchrotron emissivity at some radio frequency is 
$q\rs{sych}\propto KB^{(s+1)/2}$, $B$ is the strength of magnetic field. 
The $\gamma$-ray emissivity of electrons due to inverse Compton process is calculated as 
\begin{equation}
 q\rs{ic}(\varepsilon)=\int_{0}^{\infty}N(E)p\rs{ic}(E,\varepsilon)dE  %erg/s/cm^3/Hz;    [N(E)]=cm^-3/eV, [dE]=eV
 \label{IC-emiss}
\end{equation}
where $\varepsilon$ is the photon energy. 
The spectral distribution $p\rs{ic}$ of radiation power of a "single" electron in a black-body 
photon field with temperature $T$ is 
\begin{equation}
 p\rs{ic}(\gamma,\varepsilon)= \frac{2e^4 \epsilon\rs{c}}{\pi \hbar^3c^2} \gamma^{-2}
 {\cal I}\rs{ic}(\eta\rs{c},\eta\rs{o})
\end{equation} 
where $\gamma$ is Lorenz factor of electron, 
$\epsilon\rs{c}=kT$, 
\begin{equation}
 \eta\rs{c}={\epsilon\rs{c}\varepsilon\over \left(m\rs{e}c^2\right)^2}, \quad 
 \eta\rs{o}={\varepsilon^2\over 4\gamma m\rs{e}c^2(\gamma m\rs{e}c^2-\varepsilon)},
\end{equation}
$m\rs{e}$, $e$, $c$, $\hbar$, $k$ have their typical meaning.
${\cal I}\rs{ic}(\eta\rs{c},\eta\rs{o})$ may be approximated as \citep{Pet08IC}
\begin{equation}
 \begin{array}{ll}
 {\cal I}\rs{ic}(\eta\rs{c},\eta\rs{o})&\approx 
 \displaystyle
 \frac{\pi^2}{6}\eta\rs{c} \left(
 \exp\left[-\frac{5}{4}\left(\frac{\eta\rs{o}}{\eta\rs{c}}\right)^{1/2}\right] \right.\\
 &\displaystyle
 \left.+2\eta\rs{o}
 \exp\left[-\frac{5}{7}\left(\frac{\eta\rs{o}}{\eta\rs{c}}\right)^{0.7}\right]
 \right)
 \exp\left[-\frac{2\eta\rs{o}}{3\eta\rs{c}}\right].
 \end{array}
 \label{calIappranyeta}
\end{equation}
This approximation is quite accurate, it represents ${\cal I}\rs{ic}$ in any regime, from Thomson to extreme Klein-Nishina. The maximum of spectral distribution $p\rs{ic}(\varepsilon)$ for electrons with energy $E$ 
is at \citep{Pet08IC}
\begin{equation}
 \varepsilon\rs{max}(E)\approx\frac{E\Gamma\rs{c}}{1+\Gamma\rs{c}},\quad
 \Gamma\rs{c}=\frac{4\epsilon\rs{c}E}{(m\rs{e}c^2)^2}.
\end{equation}
All IC images in the present paper (except of that on Fig.~\ref{thetaK:fig6}) are calculated for 
the initial photon field with $T=2.75$ K and for 
the \g-ray photon energy $\varepsilon=0.1\varepsilon\rs{max}(E\rs{max})$ that is for example $\varepsilon=0.3\un{TeV}$ for $E\rs{max}=30\un{TeV}$. 

%%%%%%%%%%%%%%%%%%%%%%%%%%%%%%%%%%%%%%%%%%%%%%%%%%%%%%%%%%%%%%%%%%%%%%%%%%%%%%%%%%
\section{Results}

\subsection{Synchrotron radio images}

We stress that all figures in the present paper 
have been computed using complete MHD model. 

Let us define an aspect angle $\phi\rs{o}$ as an angle between interstellar
magnetic field and the line of sight (Fig.~\ref{thetaK:angles}). 
It is shown that the azimuthal variation of the 
radio surface brightness $S\rs{\varrho}$ at a given radius of projection $\varrho$, 
in SNR which is not centrally brightened, 
is mostly determined by the variations of the magnetic field compression 
(and/or amplification) $\sigma\rs{B}$ and the electron injection efficiency $\varsigma$ 
\citep{pet-SN1006mf}: 
\begin{equation}
 S\rs{\varrho}(\varphi)\propto 
 \varsigma\big(\Theta\rs{o,eff}(\varphi,\phi\rs{o})\big)\ 
 \sigma\rs{B}\big(\Theta\rs{o,eff}(\varphi,\phi\rs{o})\big)^{(s+1)/2}
 \label{ISMF:azimuthal}
\end{equation}
where $\varphi$ is the azimuthal angle. 
The effective obliquity angle $\Theta\rs{o,eff}$ is related to $\varphi$ and $\phi\rs{o}$ as
\begin{equation}
 \cos\Theta\rs{o,eff}\left(\varphi,\phi\rs{o}\right)=\cos\varphi\sin\phi\rs{o},
\end{equation}
here, the azimuth angle $\varphi$ is measured from the direction of ISMF in the
plane of the sky (Fig.~\ref{thetaK:angles}).

\begin{figure}
 \centering
 \includegraphics[width=8.0truecm]{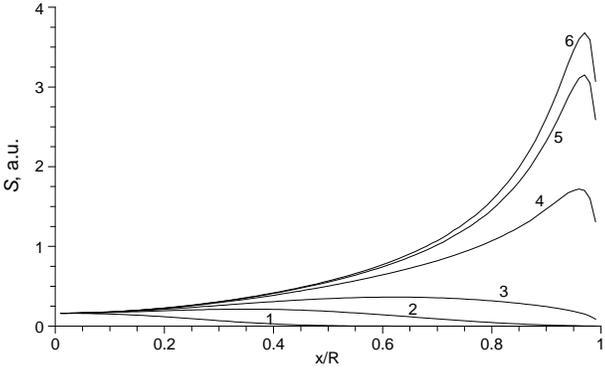}
 \caption{Profiles of the radio surface brightness for 
 an aspect angle $\phi\rs{o}=0^\mathrm{o}$ 
 (the radial profile of brightness is the same for any azimuth). 
 $\Theta\rs{K}$ is 
 $\pi/12$ (line 1), $\pi/6$ (line 2), $\pi/4$ (line 3), 
 $\pi/2$ (line 4), $\pi$ (line 5), $2\pi$ (line 6). 
                }
 \label{thetaK:fig2}
\end{figure}

Fig.~\ref{thetaK:fig1} 
shows how $\Theta\rs{K}$ affects a radio image of SNR. Complete MHD simulations are  
in agreement with the approximate formula (\ref{ISMF:azimuthal}). First, we note 
that {\em smooth increase of $\Theta\rs{K}$ results in transition from the 3-D polar-cap model of SNR 
to the 3-D barrel-like one}. 
This is also visible on Fig.~\ref{thetaK:fig2} where ISMF is directed toward observer. Namely, increase of $\Theta\rs{K}$ change the visual morphology 
from centrally-bright to shell-like. 

%--------------------------------------------------------------
%--------------------------------------------------------------

There are three names for a class of SNRs which have two opposite limbs in the literature: 
`barrel-shaped' \citep{kesteven-caswell-1987},  
`bipolar' \citep{reyn-fulbr-90} and `bilateral' \citep{gaensler-1998}. 
They were introduced on the base of 2-D visual morphology. 
It is interesting that the first two names reflects de facto the two different conceptions of SNRs in 3-D. 

Fig.~\ref{thetaK:fig1} also shows that an assumption about orientation of ISMF leads 
to limitation of possible injection model. 
Ambient magnetic field in all images on Fig.~\ref{thetaK:fig1} is along horizontal axis. 
Thus, {\em if one consider the polar-cap scenario for bilateral SNR} 
(ISMF is along axis which crosses two limbs) {\em then one should consider the 
injection model which strongly depends on the obliquity} ($\Theta\rs{K}\leq \pi/6$, 
Fig.~\ref{thetaK:fig1}a,b). 
Instead, {\em if the barrel is the preferable model} (ISMF is 
parallel to the symmetry axis between two limbs) {\em then the injection efficiency should be almost 
independent of obliquity} ($\Theta\rs{o}\geq \pi$, Fig.~\ref{thetaK:fig1}e,f), or prefer quasiperpendicular shocks. 

\citet{gaensler-1998} measured the angle $\psi$ between the symmetry axis in 
17 `clearly' bilateral SNRs and the Galactic plane. Axes are more or less aligned with the Galactic plane in 12 SNRs ($\psi<30^\mathrm{o}$), 2 SNRs have $\psi\approx 45^\mathrm{o}$ and 3 SNRs is almost perpendicular 
($\psi>60^\mathrm{o}$). If we assume that ISMF is parallel to the plane of Galaxy then most of bilateral SNRs should be 
3-D barrels preffering thus isotropic (or quasiperpendicular) injection. 

An interesting feature appears on images for $\Theta\rs{K}=\pi/4\div\pi/2$ (Fig.~\ref{thetaK:fig1}c,d). 
Namely, SNR has `quadrilateral' morphology. 
With increasing of obliquity, the injection efficiency decreases while the compression factor of ISMF icreases. 
The variation of injection $\varsigma(\Theta\rs{o})$ dominates $\sigma\rs{B}(\Theta\rs{o})$ for $\Theta\rs{K}\leq\pi/6$. If $\Theta\rs{K}\geq\pi$ (injection is almost isotropic) then $\sigma\rs{B}(\Theta\rs{o})$ plays the main role in azimuthal variation of the radio surface brightness. In the intermediate range of $\Theta\rs{K}$, the significance of the two variations are comparable leading therefore to azimuthal migration of the brightness maxima in the modelled images. 
There is no `quadrilateral' SNR reported in the literature. If there is no such SNR at all, 
the range $\Theta\rs{K}\simeq \pi/4\div\pi/2$ may be excluded. 
However, we stress that a complete statistical study of the morphology of radio SNRs 
would be needed to definitly asses the lack of quadrilateral SNRs\footnote{G338.3-0.0 could be an example of quadrilateral SNR}. 

\begin{figure*}
 \centering
 \includegraphics[width=17truecm]{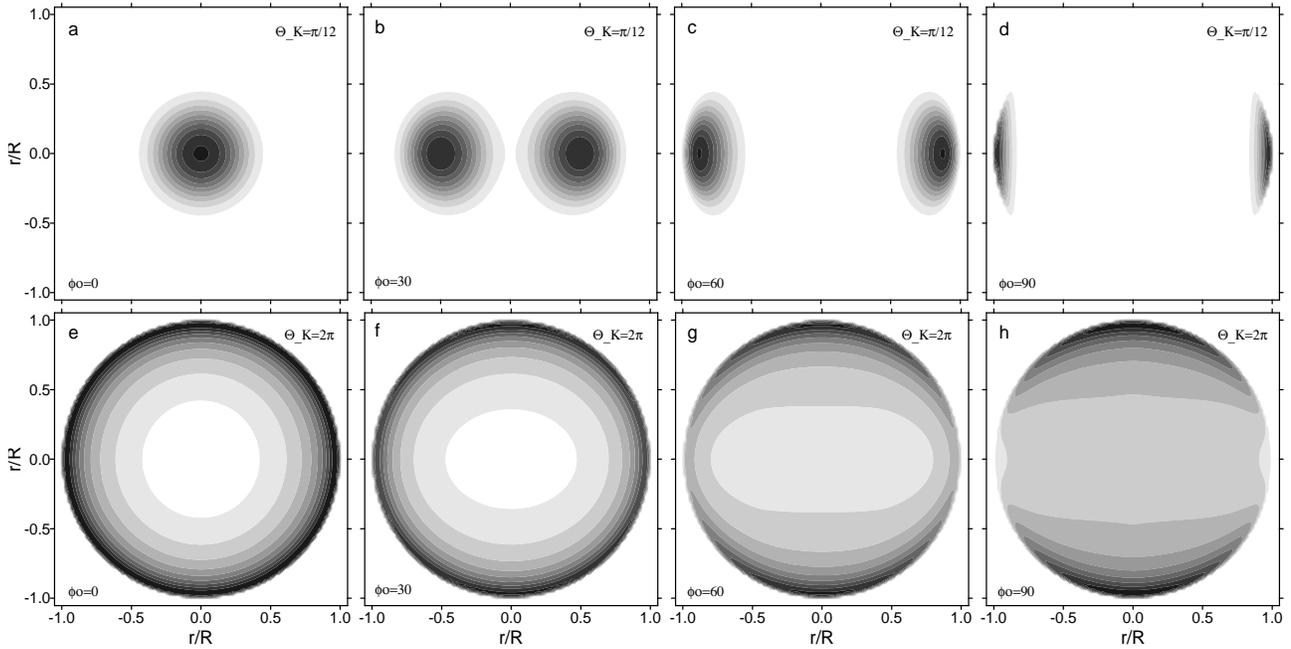}
 \caption{Radio images of SNR for different aspect angles $\phi\rs{o}$: 
 $0^\mathrm{o}$ (a,e), $30^\mathrm{o}$ (b,f), $60^\mathrm{o}$ (c,g), $90^\mathrm{o}$ (d,h). 
 $\Theta\rs{K}=\pi/12$ (upper panel), $\Theta\rs{K}=2\pi$ (lower panel).
 Component of the ambient magnetic field which is perpendicular to the line of sight, 
 is oriented along the horizontal axis. 
% Increment in brightness is $\Delta S=0.1 S\rs{\max}$.
                }
 \label{thetaK:fig3}
\end{figure*}

The visual morphology of SNR is different for different aspect angles. Fig.~\ref{thetaK:fig3} shows SNR images for quasi-parallel injection with $\Theta\rs{K}=\pi/12$ (upper panel) and for isotropic injection ($\Theta\rs{K}=2\pi$, lower panel). 
We may expect that ISMF may have different orientation versus observer in various SNRs. If quasi-parallel injection is not a rare exception then the polar-cap SNRs should be projected in a different way and we may expect to observe not only `bipolar' SNRs (Fig.~\ref{thetaK:fig3}c,d) but also SNRs with one or two radio eyes within thermal X-ray rim (Fig.~\ref{thetaK:fig3}a,b). \citet{reyn-fulbr-90} developed statistically this thought and showed that the quasi-parallel injection model would be unlikely, but again, we would need a complete study to verify this statement\footnote{G311.5-0.3 and G337.2-0.7 could be examples of SNRs with two radio 'eyes'}. 
Statistical arguments of \citet{reyn-fulbr-90} may be affected by the fact that 
centrally-bright radio SNRs (lines 1-2 on Fig.~\ref{thetaK:fig2}) are expected to be fainter than bilateral or circular SNRs with the same characteristics (lines 4-6 on Fig.~\ref{thetaK:fig2}): it could be that most of the centrally-peaked SNRs may not be observable.

%--------------------------------------------------------------
%--------------------------------------------------------------
%--------------------------------------------------------------
\begin{figure*}
 \centering
 \includegraphics[width=17truecm]{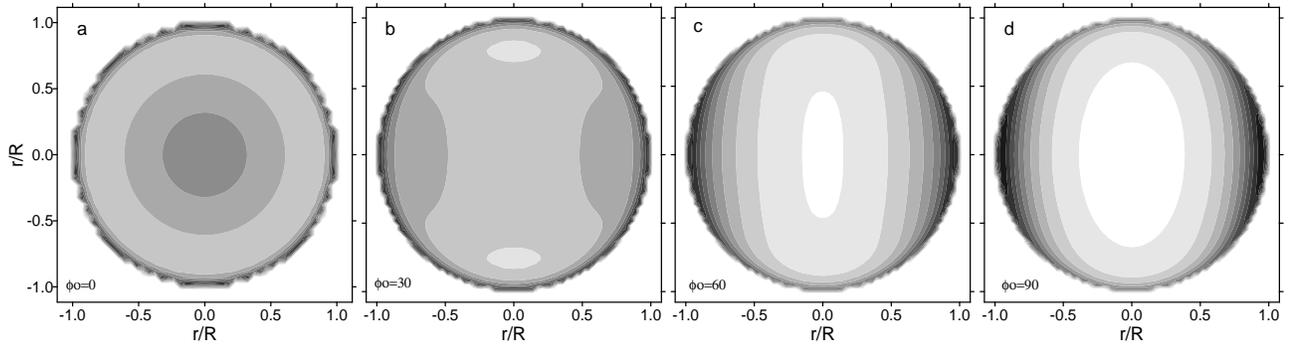}
 \caption{IC \g-ray images of SNR. Isotropic injection, $E\rs{max}$ is constant over SNR surface. 
 Aspect angles $\phi\rs{o}$: 
 $0^\mathrm{o}$ (a), $30^\mathrm{o}$ (b), $60^\mathrm{o}$ (c), $90^\mathrm{o}$ (d). 
 Component of the ambient magnetic field which is perpendicular to the line of sight, 
 is oriented along the horizontal axis. 
% Increment in brightness is $\Delta S=0.1 S\rs{\max}$.
                }
 \label{thetaK:fig4}
\end{figure*}
%--------------------------------------------------------------

\subsection{IC \g-ray images}

\op{Let us consider first the case when the maximum energy of electrons is constant over SNR surface; this allows us to clearly see the role of the injection efficiency and magentic field variations.} 

Synthesized IC \g-ray images of SNRs are presented on Fig.~\ref{thetaK:fig4}, for different aspect angles. These images assume almost {\em isotropic} injection ($\Theta\rs{K}=2\pi$) and should be compared with radio maps on the lower panel of  Fig.~\ref{thetaK:fig3}. The component of ISMF which is perpendicular to the line of sight is along horizontal axis on all images. An important difference is prominent from these two figures. Namely, the two lobes develop with increasing of $\phi\rs{o}$ in both radio and \g-rays. However, {\em their location in respect to ISMF is opposite}. The line conecting two maxima in radio is perpendicular to ISMF while it is parallel to ISMF on IC images (cf. Fig~\ref{thetaK:fig4}d and Fig~\ref{thetaK:fig3}h). 

The reason of this effect is the following. For assumed isotropic injection, the azimuthal variation of the radio brightness is determined only by the dependence $\sigma\rs{B}$ on obliquity (the azimuth angle equals to the obliquity angle for $\phi\rs{o}=\pi/2$). 
Electrons emitting VHE \g-rays have energies $E\sim E_{\max}$ and experience substantial radiative losses (this effect is negligible for radio emitting electrons). 
Magnetic field does not appear directly in the formulae for IC emission, but it affects the downstream distribution of relativistic electrons emitting IC \g-rays. The larger post-shock magnetic field the larger radiative losses. The downstream distribution of IC-emitting electrons is therefore steeper where magnetic field is stronger. This leads to lower IC brightness in SNR regions with larger magnetic field (while radio brightness increases there because of  proportionality to $B^{3/2}$). 

In VHE \g-ray image of SN~1006 recently reported by H.E.S.S. collaboration (H.E.S.S. Source of the Month, August 2008), 
the two maxima coincide in location with 
limbs in radio and nonthermal X-rays. This fact, in view of the `limb-inverse' property, could be considered as argument against the leptonic origin of \g-ray emission in SN~1006 (if injection is isotropic). However, these IC images are obtained under assumption that $E\rs{max}$ does not vary over SNR surface. If $E\rs{max}$ is high enough at regions with large magnetic field (at perpendicular shock), then the `limb-inverse' effect may be less prominent or even might not be important (see below). 

\begin{figure}
 \centering
 \includegraphics[width=8.4truecm]{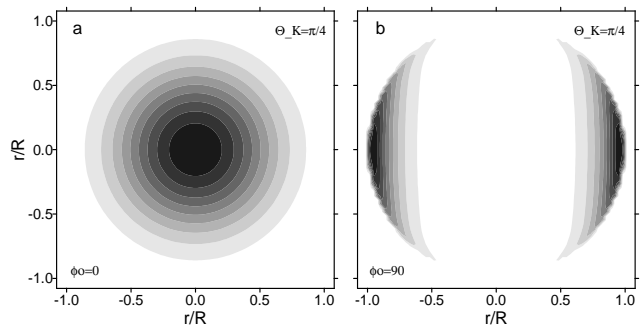}
 \caption{IC \g-ray images of SNR. Quasi-parallel injection (\ref{finj}) 
 with $\Theta\rs{K}=\pi/4$, $E\rs{max}(\Theta\rs{o})=\mathrm{const}$. Aspect angles $\phi\rs{o}$: 
 $0^\mathrm{o}$ (a), $90^\mathrm{o}$ (b). In the latter, ISMF is along the horizontal axis. 
 % (to be compared with Fig.~\ref{thetaK:fig3}a,d).
% Increment in brightness is $\Delta S=0.1 S\rs{\max}$.
                }
 \label{thetaK:fig5}
\end{figure}

In case if injection strongly prefers {\em parallel} shocks (limbs in SN~1006 are polar caps), the dependence $\varsigma(\Theta\rs{o})$ might dominate $\sigma\rs{B}(\Theta\rs{o})$. The maxima of brightness in radio and IC \g-rays are therefore located at the same regions of SNR projection (Fig.~\ref{thetaK:fig5}, to be compared with Fig.~\ref{thetaK:fig3}a,d), in agreement with the Chandra and H.E.S.S. observations of SN~1006. 

\begin{figure}
 \centering
 \includegraphics[width=8.0truecm]{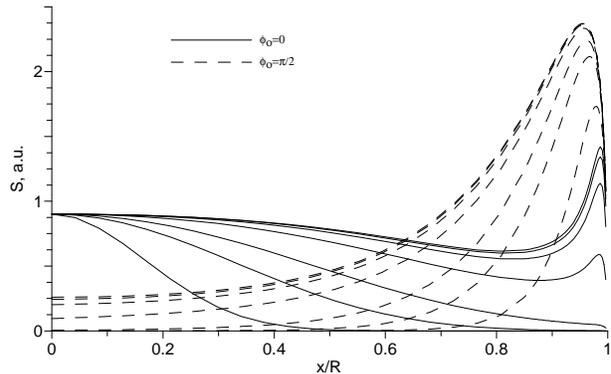}
 \caption{Profiles of the IC surface brightness along X-axis for 
 the aspect angle $\phi\rs{o}=0^\mathrm{o}$ (the radial profile 
 of brightness is the same for any azimuth; to be compared with 
 Fig.~\ref{thetaK:fig2}) and $\phi\rs{o}=90^\mathrm{o}$ 
 (ISMF is along the horizontal axis). 
 Dependence of injection is given by (\ref{finj}) with 
 $\Theta\rs{K}$ (from below):
 $\pi/12$, $\pi/6$, $\pi/4$, $\pi/2$, $\pi$, $2\pi$, $\infty$. 
 $E\rs{max}$ is constant over SNR surface.
                }
 \label{thetaK:fig8}
\end{figure}

The role of intermediate values $\Theta\rs{K}$ for injection which prefers parallel shock, 
Eq.~(\ref{finj}), on profiles of IC brightness is shown on Fig.~\ref{thetaK:fig8}. Increase of 
the sencitivity of injection to the obliquity leads to radially thinner and more contrast features. 

\begin{figure}
 \centering
 \includegraphics[width=8.4truecm]{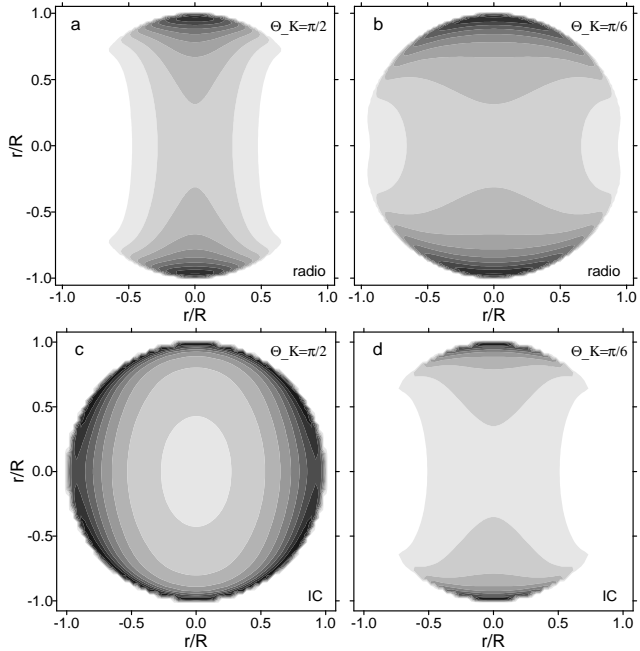}
 \caption{Radio (a,b) and IC \g-ray images (c,d) of SNR for 
 $\phi\rs{o}=90^\mathrm{o}$. Quasi-perpendicular 
 injection (\ref{finjperp}) with $\Theta\rs{K}=\pi/2$ (a,c) 
 and $\Theta\rs{K}=\pi/6$ (b,d) 
 (to be compared with Fig.~\ref{thetaK:fig3}d and 
 Fig.~\ref{thetaK:fig4}d). $E\rs{max}$ is constant over SNR surface. 
                }
 \label{thetaK:fig7}
\end{figure}

If injection prefers {\em perpendicular} shock, Eq.~(\ref{finjperp}), its 
increase in the regions of larger magnetic field may compensate the lack 
of \g-ray emitting electrons. 
In that case, the position of limbs coincide in radio and IC \g-rays 
if the dependence $\varsigma(\Theta\rs{o})$ 
is strong enough (Fig.~\ref{thetaK:fig7}b,d). In the range of intermediate $\Theta\rs{K}$, 
the quadrilateral morphology appears also in models of IC \g-rays 
(Fig.~\ref{thetaK:fig7}c), as an intermediate morphology between 
those on Fig.~\ref{thetaK:fig4}d and Fig.~\ref{thetaK:fig7}d. 
(The contrast of maxima in the image of quadrilateral SNR is so small that 
this feature may probably not be observable.)

Note that the quasi-perpendicular injection model leads to {\em radio} images similar to those in the isotropic injection case, cf. Fig.~\ref{thetaK:fig7}a,b and Fig.~\ref{thetaK:fig1}f (see also \citet{Orletal07}), because magnetic field and injection efficiency increase at perpendicular shocks both leading to larger 
synchrotron emission. In contrast, there is a lack of IC radiating electrons around perpendicular shocks which may or may not (depending on $\Theta\rs{K}$ in (\ref{finjperp})) compensate it. Thus {\em IC} images involving the quasi-perpendicular injection  may radically differ from those with isotropic injection, cf. Fig.~\ref{thetaK:fig7}d and Fig.~\ref{thetaK:fig4}d.

%--------------------------------------------------------------
%--------------------------------------------------------------

\begin{figure}
 \centering
 \includegraphics[width=8.4truecm]{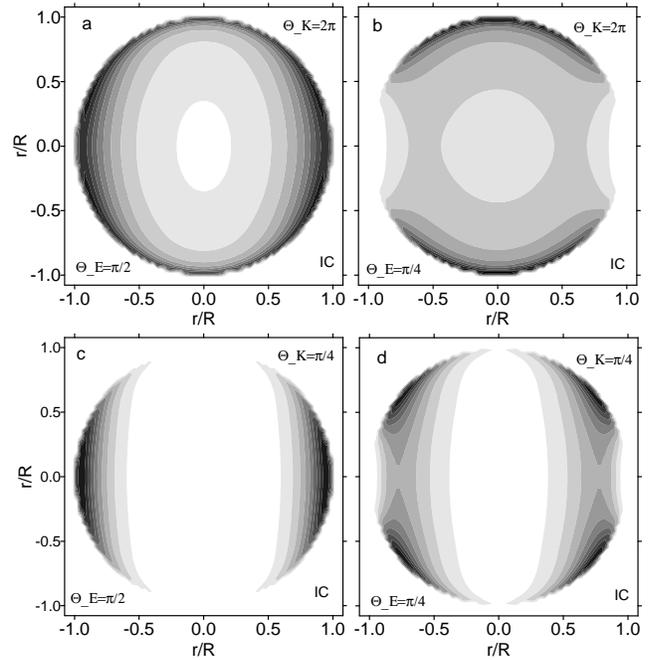}
% \vspace{4truecm}
 \caption{IC \g-ray images of SNR for 
 $\phi\rs{o}=90^\mathrm{o}$ and $E\rs{max}$ increasing with obliquity, 
 Eq.~(\ref{fEmaxperp}) with $\Theta\rs{E}=\pi/2$ (a,c) and $\Theta\rs{E}=\pi/4$ (b,d). 
 Isotropic injection (a,b), to be compared with Fig.~\ref{thetaK:fig4}d; 
 quasi-parallel injection with $\Theta\rs{K}=\pi/4$ (c,d), 
 to be compared with Fig.~\ref{thetaK:fig5}b. 
% Quasi-perpendicular injection (\ref{finjperp}) with $\Theta\rs{K}=\pi/6$ 
% and $\Theta\rs{E}=...$ (d) 
% (to be compared with Fig.~\ref{thetaK:fig7}d).
                }
 \label{thetaK:fig9}
\end{figure}

\op{
The obliquity variation of the electron maximum energy is an additional factor affecting the IC \g-ray 
brightness in SNRs. 
Actually, \citet{Rotetal04} have shown that the cut-off frequency 
increases at radio limbs of SN~1006 that may (partially) be due to larger $E\rs{max}$ there. 
Therefore $E\rs{max}$ is expected to be largest in this SNR at the perpendicular shock (at equatorial belt) if injection is isotropic or quasi-perpendicular or at the parallel shock (at polar caps) if injection is quasi-parallel.
In the latter case, the calculations of \citet{Reyn-98} suggest that the only possible model for $E\rs{max}$ in SN~1006 should be loss-limited one in the Bohm limit. 

The role of $E\rs{max}$ increasing with obliquity, Eq.~(\ref{fEmaxperp}), is shown on Fig.~\ref{thetaK:fig9}. The `limb-inverse' property may not be important and the limbs may coincide in radio, X-rays and IC \g-rays even for the isotropic injection if the maximum energy is large enough at perpendicular shocks to provide energetical electrons in  despite of radiative losses (Fig.~\ref{thetaK:fig9}b, cf. with Fig.~\ref{thetaK:fig3}h and Fig.~\ref{thetaK:fig4}d). 
Note also that the limbs are thicker in this case, because of the more effective radiative losses at perpendicular shock (due to larger ISMF compression), comparing to limbs if they are at parallel shock. 

The dependence of $E\rs{max}$ on $\Theta\rs{o}$ may also cause splitting and rotation of IC limbs in case of the quasi-parallel injection (Fig.~\ref{thetaK:fig9}d, cf. with Fig.~\ref{thetaK:fig5}b) or the quasi-perpendicular one. 
There is a possibility for quadrilateral SNRs to appear in \g-rays due to the interplay between dependences  
$E\rs{max}(\Theta\rs{o})$, $\varsigma(\Theta\rs{o})$ and $\sigma\rs{B}(\Theta\rs{o})$ (Fig.~\ref{thetaK:fig9}a,d). 
}

\begin{figure}
 \centering
 \includegraphics[width=5.1truecm]{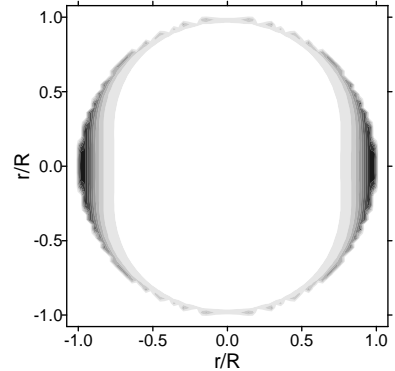}
 \caption{The same as Fig.~\ref{thetaK:fig4}d, for 10 times larger photon energy, 
 $\varepsilon=\varepsilon\rs{max}(E\rs{max})$.
                }
 \label{thetaK:fig6}
\end{figure}

All above IC images are calculated for the photon energy $\varepsilon=0.1\varepsilon\rs{max}(E\rs{max})$. 
The pattern of the \g-ray surface brightness remain 
almost the same with increasing of the photon energy, though regions of maximum brightness 
become radially thinner and also contrasts change (Fig.~\ref{thetaK:fig6}). 
This is because electrons which contribute most of emission at larger photon energy 
experience higher radiative losses and therefore the downstream distribution of these electrons are steeper.

To the end, 
the main properties of IC surface brightness may simply be derived from the approximate analytical 
formula for the azimuthal variation of IC surface brightness $S\rs{\varrho}(\varphi;\phi\rs{o},\varepsilon)$ 
of the adiabatic SNR in uniform ISM and uniform ISMF (Appendix): 
\begin{equation}
 S\rs{\varrho}(\varphi)\propto
 \displaystyle
 \varsigma(\Theta\rs{o,eff})
 \exp\left(-%\frac{E\rs{m}}{E\rs{max,\|}} 
 \frac{E\rs{m}\bar \varrho^{-1-{5\sigma\rs{B}(\Theta\rs{o,eff})^2E\rs{m}/2E\rs{f,\|}}}}
 {E\rs{max,\|}{\cal F}(\Theta\rs{o,eff})}\right)
 \label{ICazimuth:text}
\end{equation} 
where $E\rs{m}\propto \varepsilon^{1/2}$, Eq.~(\ref{ICimages:Em}), 
$\bar \varrho=\varrho/R\leq1$, $\varrho$ is the distance from the center of SNR projection.
This formula may not be used for SNR which is centrally-bright in \g-rays and is valid 
for $\varrho/R$ larger than $\simeq 0.9$.

%--------------------------------------------------------------
%--------------------------------------------------------------

%%%%%%%%%%%%%%%%%%%%%%%%%%%%%%%%%%%%%%%%%%%%%%%%%%%%%%%%%%%%%%%%%%%%%%%%%
\section{Conclusions}

In the present paper, we analyse the synchrotron radio and the inverse-Compton \g-ray images of Sedov SNRs synthesized  on the base of the \citet{Reyn-98} model. \citet{ell-bar-jones-95} have shown that the dependence of efficiency of injection $\varsigma$ on obliquity angle $\Theta\rs{o}$ may differ from commonly used expression in quasi-parellel case. We therefore parameterise the dependence $\varsigma(\Theta\rs{o})$ as it is given by Eq.~(\ref{finj}). 
It is shown that the variation of the parameter $\Theta\rs{K}$ provide smooth transition from polar-cap ($\Theta\rs{K}\leq \pi/6$) to barrel-like ($\Theta\rs{K}\geq \pi$) models of SNR and that assumed orientation of ISMF should be related to a certain injection model. Some constraints on injection models which follow from morphological considerations are pointed out. The azimuthal variation of radio brightness is mostly due to variations of $\varsigma$ and $\sigma\rs{B}$, in agreement with the approximate formula (\ref{ISMF:azimuthal}).

Theoretical \g-ray images of SNR due to the inverse Compton effect are reported for the first time. We analyse  properties of these images and compare them with corresponding radio maps of SNRs. 
The azimuthal variation of IC brightness is mostly determined by variations of $\varsigma$, $\sigma\rs{B}$ and $E\rs{max}$, in agreement with the approximate formula (\ref{ICazimuth:text}) derived in the Appendix. 

\op{In case if $E\rs{max}$ is constant over the SNR surface,} we found an opposite behaviour of azimuthal variation of surface brightness in radio and IC \g-rays, in case if injection is isotropic and the aspect angle is larger than $\simeq 60^\mathrm{o}$. 
Namely, the line crossing the two limbs in radio are perpendicular to the ISMF while 
they are parallel in IC \g-rays. 
In particular, bright radio limbs correspond to dark IC areas, in disagreement with X-ray and H.E.S.S. observations of SN~1006. 
This happens because IC image is affected by large radiative losses of emitting electrons behind perpendicular shock while the larger magnetic field increases the radio brightness there. Variation of $E\rs{max}$ over SNR surface may (to some extent) hide this effect. The maximum energy should increase with obliquity in this case. 

In case of the polar-cap model of SNR (quasi-parallel injection), the maxima in surface brightness are expected to coincide in radio and IC \g-rays (in agreement with H.E.S.S. observation of SN~1006), 
\op{unless increase of $E\rs{max}$ with obliquity will be very strong, which is unlikely 
in case of SN~1006 because the cut-off frequency is larger at limbs 
which are at parallel shock in this injection model.} 

Limbs may also coincide in case of the quasi-perpendicular injection, 
if the lack of electrons (due to radiative losses) in the regions of large magnetic field 
is compensated by the strong enough increase of $\varsigma$ \op{and/or $E\rs{max}$} with $\Theta\rs{o}$. 

Isotropic compression/amplification of ISMF on the shock (i.e. independent of the shock obliquity), like it could be under highly effective acceleration, may also be responsible for the same position of limbs in radio and in IC \g-rays, for the quasi-parallel or  quasi-perpendicular injection scenarios. \op{In this case the dependence of $E\rs{max}(\Theta\rs{o})$ have to follow variation $\varsigma(\Theta\rs{o})$, namely, to be largest (smallest) at parallel shock for quasi-parallel (quasi-perpendicular) injection, otherwise the morphology of SNR in IC \g-rays may differ from the radio one.}

%--------------------------------------------------------------
%--------------------------------------------------------------
%--------------------------------------------------------------
%--------------------------------------------------------------
%--------------------------------------------------------------
%--------------------------------------------------------------

\op{We conclude that the location the \g-ray limbs versus radio and X-ray ones, 
recently discovered by H.E.S.S. in SN~1006, cannot be 
conclusive about the actual dependence of the 
electron injection efficiency, the compression/amplification of ISMF and 
the electron maximum energy on the obliquity angle in this SNR. 
Detailed features of the SNR maps in different wavebands should be considered for this purpose. }

\op{The interplay between dependences $\varsigma(\Theta\rs{o})$, $\sigma\rs{B}(\Theta\rs{o})$ and $E\rs{max}(\Theta\rs{o})$ may cause the quadrilateral morphology in SNR models, due to splitting of maxima in surface brightness. Absence of quadrilateral SNRs in IC \g-rays, if revealed observationally, may results in limitations on $\Theta\rs{K}$ and $\Theta\rs{E}$.} 

%\op{Absence of quadrilateral SNRs in IC \g-rays, if revealed, put limitations on $\Theta\rs{K}$, $\Theta\rs{K}$. 
%and may results in conclusion that functions 
%$\varsigma(\Theta\rs{o})$ and $E\rs{max}(\Theta\rs{o})$ vary in the same 
%way (both increase or decrease with obliquity), or some of them may be constant.}

The detailed characterictics of features on IC image (e.g. thickness of rim) depend on the photon energy. They are radially thinner at larger photon energies, as expected. 

\section*{Acknowledgments}

OP acknowledge Osservatorio Astronomico di Palermo for hospitality. 
The work of OP was partially supported by 
the program 'Kosmomikrofizyka' of National Academy of Sciences (Ukraine). 
FB, SO and OP acknowledge Consorzio COMETA under the PI2S2 Project, a
project co-funded by the Italian Ministry of University and Research
(MIUR) within the Piano Operativo Nazionale `'Ricerca Scientifica,
Sviluppo Tecnologico, Alta Formazione' (PON 2000-2006).

%%%%%%%%%%%%%%%%%%%%%%%%%%%%%%%%%%%%%%%%%%%%%%%%%%%%%%%%%%%%%%%%%%%%%%%%%

%%%%%%%%%%%%%%%%%%%%%%%%%%%%%%%%%%%%%%%%%%%%%%%
\appendix
\section[]{Approximate analytical formula for the azimuthal variation of the IC \g-ray surface brightness in Sedov SNR}

An approximate formula for azimuthal variation of the IC \g-ray surface brightness 
allows one to avoid detailed numerical simulations and may 
be useful if approximate estimation for the variation is reasonable. 
It gives deeper insight in the main factors
determining the azimuthal behavior of the IC surface brightness in SNRs. 

Let the energy of relativistic electrons is $E$ in a given fluid element at present time. 
Their energy was $E\rs{i}$ at the time this element was shocked. 
These two energies are related as 
\begin{equation}
 E=E\rs{i}{\cal E}\rs{ad}{\cal E}\rs{rad}
\end{equation}
where ${\cal E}\rs{ad}$ accounts for the adiabatic losses and ${\cal E}\rs{rad}$ for the radiative losses. 
There are approximations valid close to the shock \citep{Pet-Beshl-en-2008}: 
\begin{equation} 
 {\cal E}\rs{ad}\approx \bar a,\qquad {\cal E}\rs{rad}\approx \bar a^{5\sigma\rs{B}^2E/2E\rs{f,\|}}
 \label{approxE}
\end{equation}
where $\bar a=a/R$, $a$ is Lagrangian coordinate of the fluid element, 
$E\rs{f,\|}$ is the fiducial energy for parallel shock. 
The downstream evolution of $K$ in a Sedov SNR is 
\begin{equation}
 K\propto\varsigma(\Theta\rs{o})\bar K(\bar a).
\end{equation}
With the approximations (\ref{approxE}), the distribution  
$N(E)$ may be written in the model of \citet{Reyn-98} as
\begin{equation}
 N(E,\Theta\rs{o})\propto \varsigma(\Theta\rs{o})\bar K(\bar a)E^{-s}
 \exp\left(-\frac{E\bar a^{-\psi(E,\Theta\rs{o})}}{E\rs{max,\|}{\cal F}(\Theta\rs{o})}\right)
\end{equation}
where 
\begin{equation}
 \psi(E,\Theta\rs{o})=1+\frac{5\sigma\rs{B}(\Theta\rs{o})^2E}{2E\rs{f,\|}}
\end{equation}
and the obliquity variation of the maximum energy of electrons is given by 
$E\rs{max}=E\rs{max,\|}{\cal F}(\Theta\rs{o})$. 

Electrons with Lorentz factor $\gamma$ emit most of their IC radiation in photons with energy $\varepsilon\rs{m}$. 
Let us use the 'delta-function approximation' \citep{Pet08IC}: 
\begin{equation} 
 p\rs{ic}(\gamma,\varepsilon)\approx p\rs{m}(\gamma)\delta(\varepsilon-\varepsilon\rs{m}),
 \quad
 p\rs{m}(\gamma)=\int\limits_{0}^{\infty}p\rs{ic}(\gamma,\varepsilon) d\varepsilon.
 \label{deltaformulae}
\end{equation}
In the Thomson limit, which is valid for SNRs in most cases, 
$\varepsilon\rs{m}(\gamma)\approx 4kT\gamma^2$ \citep{Pet08IC} and 
$p\rs{m}(\gamma)=(4/3)c\sigma\rs{T}\omega \gamma^2$ \citep{Schlick-book},  
$T$ and $\omega$ are the temperature and the energy density of initial black-body photons, 
$\sigma\rs{T}$ is the Thomson cross-section. 

Substitution (\ref{IC-emiss}) with (\ref{deltaformulae}) yields 
\begin{equation} 
 q\rs{ic}= 
 \frac{c\sigma\rs{T}\omega m\rs{e}c^2\varepsilon^{1/2}}{12\epsilon\rs{c}^{3/2}}N(E\rs{m})
\end{equation}
where 
\begin{equation}
 E\rs{m}=\frac{m\rs{e}c^2\varepsilon^{1/2}}{2(kT)^{1/2}}
 \label{ICimages:Em}
\end{equation} 
is the energy of electrons which give maximum contribution to IC emission at photons with energy $\varepsilon$. 

Let us consider the azimuthal profile of the IC \g-ray brightness $S\rs{\varrho}$ 
at a given radius $\varrho$ from the centre of the SNR projection. 

The obliquity angle $\Theta\rs{o}$ is different for each radial sector of 3-D object. 
It is determined, for any position within SNR, by the set 
$(\varphi,\bar r/\bar \varrho,\phi\rs{o})$. 
Integration along the line of sight gathers information from different radial 
sectors, with different obliquities. 
Let us determine the `effective' obliquity angle by the relation 
\begin{equation}
 \Theta\rs{o,eff}(\varphi,\phi\rs{o})=\Theta\rs{o}(\varphi,1,\phi\rs{o}). 
\end{equation}
Actually, $\Theta\rs{o,eff}$ 
for a given azimuth equals to the obliquity angle for a sector with the same azimuth 
lying in the plane of the sky (i.e. in the plane being perpendicular to the line of sight 
and containing the center of SNR). 
$\Theta\rs{o}$ varies around $\Theta\rs{o,eff}$ during integration along the line of sight. 
The closer $\varrho$ to the edge of SNR projection the smaller the range for 
variation of $\Theta\rs{o}$ and more accurate is our approximation. 

The surface brightness of SNR projection at distance $\varrho$ from the center and at azimuth $\varphi$ is 
\begin{equation}
 S(\bar\varrho,\varphi)=2\int^{1}_{\bar a(\bar\varrho)}q\rs{ic}(\bar a) {\bar r \bar r\rs{\bar a} d\bar a\over 
 \sqrt{\bar r^2-\bar \varrho^2}}.
\end{equation}
where $\bar r\rs{\bar a}$ is the derivative of $\bar r(\bar a)$ in respect to $\bar a$.
The azimuthal variation of the IC brightness for fixed $\varrho$  
is approximately 
\begin{equation}
\begin{array}{ll}
 S\rs{\varrho}&\propto
 \displaystyle
 \varsigma(\Theta\rs{o,eff})
 \exp\left(-\frac{E\rs{m}\bar \varrho^{-\psi\left(E\rs{m},\Theta\rs{o,eff}\right)}}
 {E\rs{max,\|}{\cal F}(\Theta\rs{o,eff})}\right)\\ \\
 &\times\displaystyle
 \int^{1}_{\bar a(\bar\varrho)}{\bar r \bar r\rs{\bar a} d\bar a\over 
 \sqrt{\bar r^2-\bar \varrho^2}} 
 \exp\left(-\frac{E\rs{m}(\bar a^{-\psi}-\bar \varrho^{-\psi})}
 {E\rs{max,\|}{\cal F}}\right)
\end{array}
 \label{ICazimuth:eq2}
\end{equation}
If $\bar \varrho\rightarrow 1$ then $\bar a(\bar\varrho)\rightarrow 1$.
Thus, the exponent in the integral is roughly unity because $\bar a(\bar\varrho)\leq\bar a\leq 1$ and $\bar a(\bar\varrho)\leq\bar \varrho\leq 1$. The integral in (\ref{ICazimuth:eq2}) is therefore roughly the same for any azimuthal
angle $\varphi$. The azimuthal variation of the IC \g-ray brightness $S\rs{\varrho}(\varphi;\phi\rs{o},\varepsilon)$ is thus determined mostly just by
\begin{equation}
 S\rs{\varrho}(\varphi)\propto
 \varsigma(\Theta\rs{o,eff})
 \exp\left(-%\frac{E\rs{m}(\varepsilon)}{E\rs{max,\|}} 
 \frac{E\rs{m}\bar \varrho^{-1-{5\sigma\rs{B}(\Theta\rs{o,eff})^2E\rs{m}/2E\rs{f,\|}}}}
 {E\rs{max,\|}{\cal F}(\Theta\rs{o,eff})}\right)
 \label{ICazimuth}
\end{equation}
with $E\rs{m}$ given by Eq.~(\ref{ICimages:Em}), i.e. $S\rs{\varrho}$ depends in this approximation 
on the temperature $T$ of the seed black-body photons and the energy $\varepsilon$ of observed \g-photons. 
The relation between the azimuthal
angle $\varphi$, the obliquity angle $\Theta\rs{o,eff}$ and the aspect angle
$\phi\rs{o}$ is as simple as
\begin{equation}
 \cos\Theta\rs{o,eff}\left(\varphi,\phi\rs{o}\right)=\cos\varphi\sin\phi\rs{o}
\end{equation}
for the azimuth angle $\varphi$ measured from the direction of ISMF in the
plane of the sky.

The approximation (\ref{ICazimuth}) may be used for $\bar\varrho$ larger than $\simeq 0.9$.
Like Eq.~(\ref{ISMF:azimuthal}), Eq.~(\ref{ICazimuth}) does not give correct profiles 
in the case of centrally-bright SNRs, i.e. when $\Theta\rs{K}\leq\pi/4$ in (\ref{finj}) and $\phi\rs{o}<30^\mathrm{o}$. 
%(see Petruk et al. \cite{pet-SN1006mf} for details). 

%%%%%%%%%%%%%%%%%%%%%%%%%%%%%%%%%%%%%%%%%%%%%%%%%%%%%%%%%%%%

\label{lastpage}

\begin{thebibliography}{99}
 \bibitem[{{Aharonian} {et al.}(2006)}]{RX1713aha2006} 
  Aharonian, F. et al., 2006, A\&A 449, 223
 \bibitem[{{Aharonian} {et al.}(2007)}]{RX1713aha2007} 
  Aharonian, F. et al., 2007, A\&A 464, 235
 \bibitem[{{Berezhko} \& {V\"olk}(2006)}]{RX1713Ber-Volk-06} 
  Berezhko, E. G., \& V\"olk, H. J. 2006, A\&A 451, 981
 \bibitem[{{Cassam-Chena\"i} {et al.}(2005)}]{Decours-2005-prof} 
  Cassam-Chena\"i G., Decourchelle A., Ballet J., Ellison D. C., 2005, A\&A 443, 955
 \bibitem[{{Ellison} {et al.}(1995)}]{ell-bar-jones-95} 
  Ellison D. C., Baring M. G., Jones F. C., 1995 \ApJ, 453, 873
 \bibitem[{{Ellison} \& {Cassam-Chena\"i}(2005)}]{Ell-Cassam2005-profiles} 
  Ellison D. \& Cassam-Chena\"i G., 2005, ApJ, 632, 920
 \bibitem[{{Gaensler}(1998)}]{gaensler-1998} 
  {Gaensler} B.~M., 1998, \ApJ, 493, 781
 \bibitem[{{Fulbright} \& {Reynolds}(1990)}]{reyn-fulbr-90} 
  {Fulbright} M.~S., \& {Reynolds} S.~P., 1990, \ApJ, 357, 591,
 \bibitem[{{Kesteven} \& {Caswell}(1987)}]{kesteven-caswell-1987} 
  {Kesteven} M.~J. \& {Caswell} J.~L., 1987, A\&A, 183, 118
% \bibitem[2004]{Lazend2004-RX1713} Lazendic J. et al. 2004, ApJ, 602, 271
 \bibitem[{{Lee} {et al.}(2008)}]{Ell2008-images} 
  Lee S.-H., Kamae T., Ellison D. C., 2008, \ApJ, 686, 325
 \bibitem[{{Orlando} {et al.}(2007)}]{Orletal07}
  {Orlando} S., {Bocchino} F., {Reale} F., {Peres} G., \& {Petruk} O., 2007, 
  A\&A, 470, 927,
 \bibitem[{{Petruk}(2006)}]{petruk2006} 
  Petruk O., 2006 A\&A, 460, 375
 \bibitem[{{Petruk}(2008)}]{Pet08IC} 
  Petruk O., 2008, astro-ph/0807.1969
 \bibitem[{{Petruk} \& {Beshley}(2008)}]{Pet-Beshl-en-2008} 
  Petruk O., Beshley V., 2008, KPCB, 24, 159
 \bibitem[{{Petruk} {et al.}(2009)}]{pet-SN1006mf} 
 Petruk O., Dubner G., Castelletti G., Iakubovskyi D., Kirsch M., Miceli M., 
 Orlando S., Telezhinsky I., 2009, MNRAS, accepted 
 \bibitem[{{Plaga}(2008)}]{Plaga2008} 
  Plaga R., 2008, New Astronomy, 13, 73
 \bibitem[{{Reynolds}(1998)}]{Reyn-98} 
  Reynolds S. P., 1998, \ApJ, 493, 375
 \bibitem[{{Rothenflug} {et al.}(2004)}]{Rotetal04} 
 {Rothenflug} R., {Ballet} J., {Dubner} G., {Giacani} E., {Decourchelle}
  A., \& {Ferrando} P., 2004, A\&A, 425, 121
 \bibitem[{{Schlickeiser}(2002)}]{Schlick-book} 
  Schlickeiser R. Cosmic Ray Astrophysics (Springer, 2002)
\end{thebibliography}
\end{document}